\documentclass[10pt,twocolumn,prl]{revtex4-1} 

\usepackage{geometry} 
\usepackage{graphicx} 

\usepackage{amsmath}
\usepackage{epstopdf}
\usepackage[caption=false]{subfig} 

\usepackage{fancyhdr} 
\begin{document}

\title{Dynamical Localization in Molecular Alignment of Kicked Quantum Rotors}

\author{A. Kamalov$^{1,2\: \ddagger}$}
\author{D.W. Broege$^{1,3}$}
\author{P.H. Bucksbaum$^{1,2,3}$}

\affiliation{$^{1}$Stanford PULSE Institute, SLAC National Accelerator Laboratory, 2575 Sand Hill Rd., Menlo Park, CA 94025, USA}
\affiliation{$^{2}$Department of Physics, Stanford University, 450 Serra Mall, Stanford, CA, 94305, USA}
\affiliation{$^{3}$Department of Applied Physics, Stanford University, 450 Serra Mall, Stanford, CA, 94305, USA}

\begin{abstract}
The periodically $\delta$-kicked quantum linear rotor is known to experience non-classical bounded energy growth due to quantum dynamical localization in angular momentum space. We study the effect of random deviations of the kick period in simulations and experiments. This breaks the energy and angular momentum localization and increases the rotational alignment, which is the analog of the onset of Anderson localization in 1-D chains.

\begin{description}
\item[PACS numbers]
\verb+33.80.-b,+ \verb+05.45.Mt,+  \verb+72.15.Rn,+ \verb+64.60.Cn+
\end{description}
\end{abstract}

\maketitle

A consequence of quantum mechanics is a set of localization mechanisms that are inexplicable in a classical picture without state quantization. Earlier studies \cite{casati, chirikov} of the periodic $\delta$-kicked rotor found strong contrast between the energy growths of the classical and quantum rotors with subsequent kicks: The classical rotor exhibited unbounded energy growth, but the quantum rotor was shown to have bounded growth despite no limitation in the excitation bandwidth. This result was explained \cite{PhysRevLett.49.509} by showing the periodically kicked rotor to be mathematically analogous to Anderson localization in the 1-D tight binding model \cite{Anderson}, leading to localization within the quantum rotor's angular momentum state space \cite{PhysRevA.86.021401}. The localization mechanism in the quantum rotor, popularly referred to as dynamical localization, is of significant interest in the quantum chaos community but has never before been experimentally studied in a true quantum linear rotor. Dynamical localization has been studied within the context of an approximately analogous system consisting of ultracold atoms subject to periodically applied potentials created by standing waves \cite{PhysRevA.45.R19}. The ultracold atom approach culminated in multiple demonstration of dynamical localization \cite{PhysRevLett.75.4598, PhysRevLett.81.1203, PhysRevLett.80.4111,  PhysRevLett.85.2741}. A previous experimental study using true quantum rotors by Zhdanovich et all \cite{PhysRevLett.109.043003} has utilized a periodic pulse train to study energy transitions across a limited number of rotation states. Zhdanovich's work demonstrated isotope selectivity which relied on localization within the unselected isotope's angular momentum states, but did not itself show localization. This report is the first study that includes experimental evidence of the dynamical localization mechanism within the true quantum rotor. 

We investigate the impact of deviation from periodic kicking on the localization of angular momentum population distributions of quantum rotors. Simulations of 64 pulse trains show that the introduction of non-uniform separation, or spacing disorder, to the pulse time separations causes the rotor to become more oriented along the $z$-axis, the axis of kicking. This corresponds to increased angular momentum in the $x-y$ plane and the delocalization of the rotor's angular momentum state population distribution. Experiments agree with our simulations. Our findings display the dynamical localization process expected in periodically kicked quantum linear rotors.

The alignment of the quantum linear rotor $^{14}$N is studied when it is subject to multiple laser-induced impulsive kicks of the sample. Molecular alignment \cite{Fried, zon} is the spatial anisotropy of the molecule's angular orientation \cite{pabstRev, averbukhReview, stapelfeldt}. Substantial literature has been published on the alignment of linear molecules after the application of a multi-pulse trains with uniform pulse separation \cite{PhysRevLett.109.043003, PhysRevA.80.063412, PhysRevLett.107.243004, schippers}, but there has been no experimental investigation of linear molecules using pulse sequences with non-uniform time separation.

\begin{figure*}[t]
	\begin{tabular}{@{}c@{}}
		\includegraphics[width=1\textwidth]{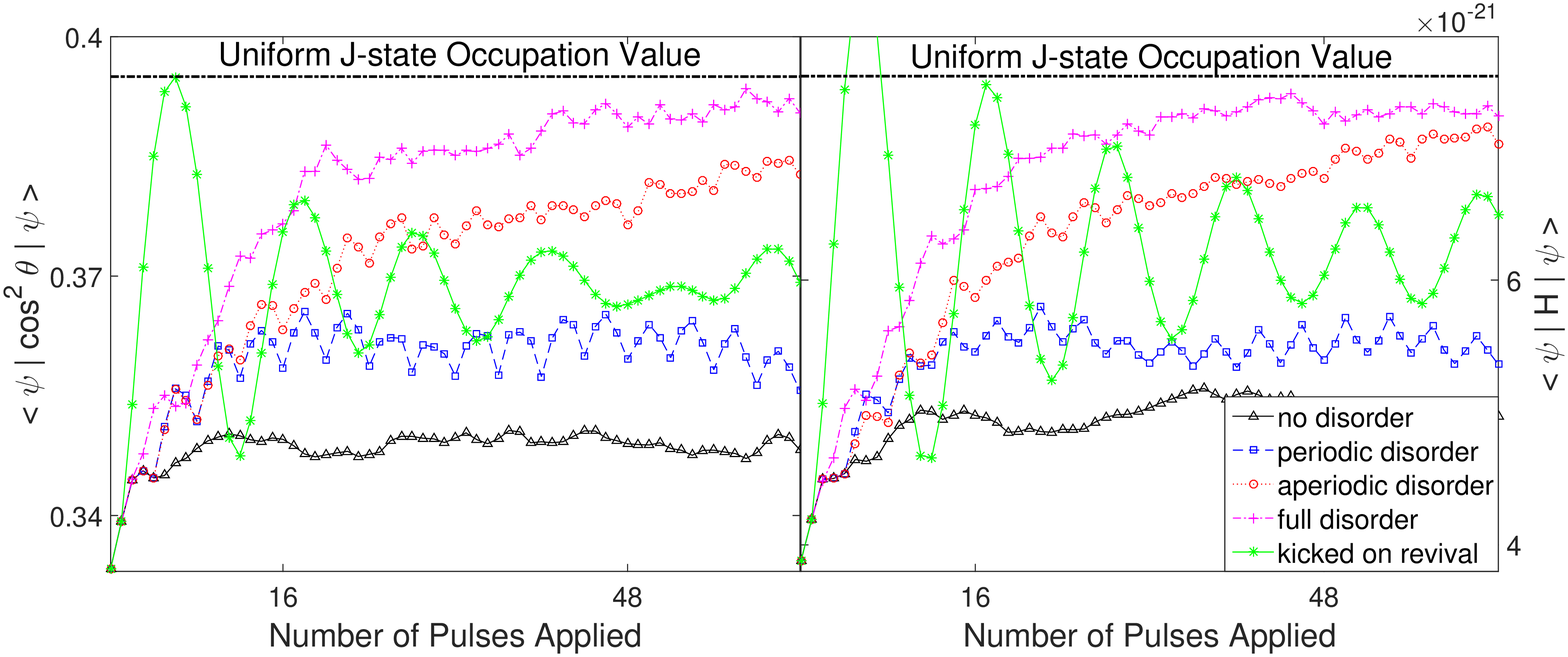} \\
		\includegraphics[width=1\textwidth]{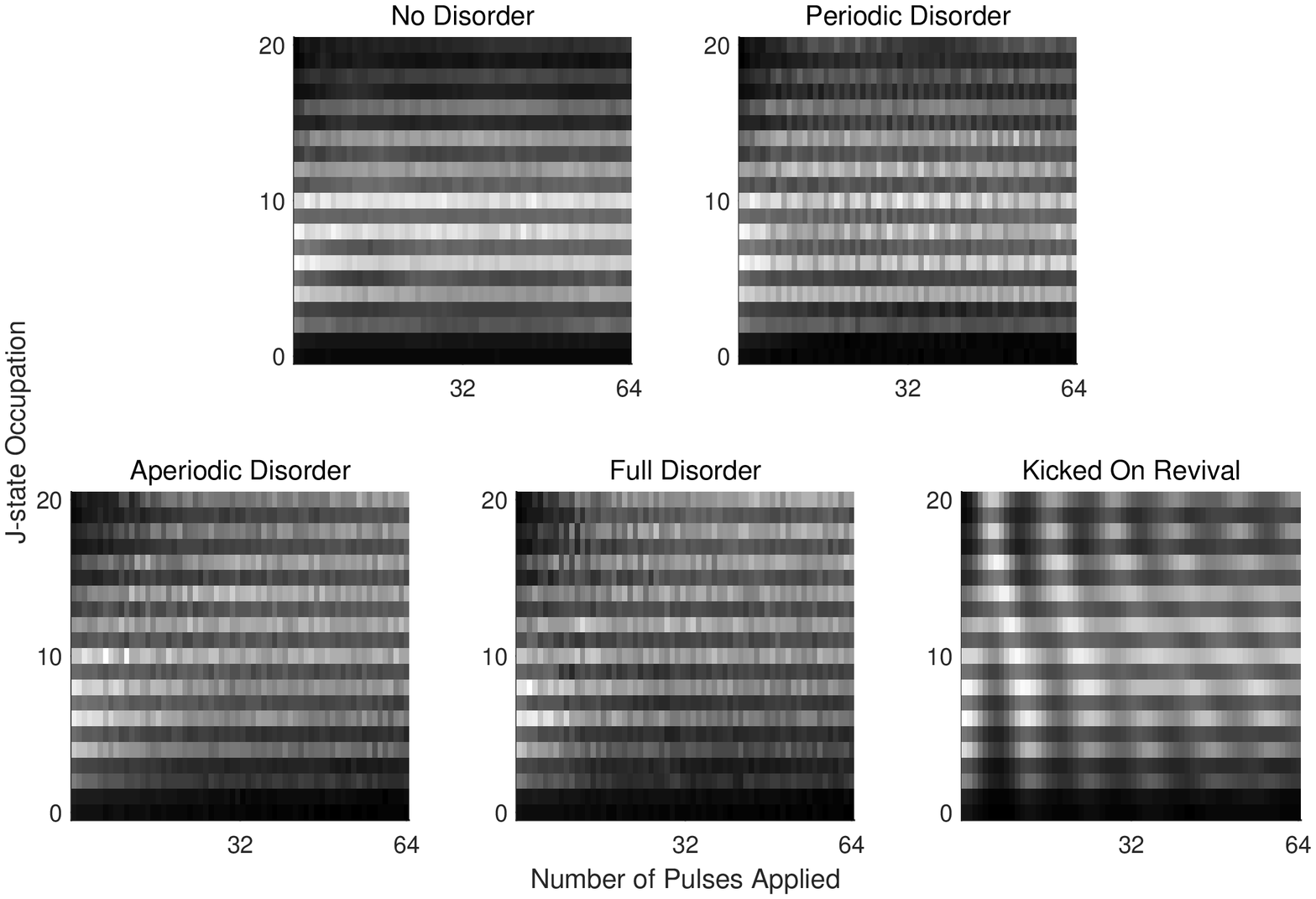}
	\end{tabular}
\caption{Simulated population alignment and energy for extended pulse trains with different levels of spacing disorder. The train with no disorder is a pulse train with uniform $\tau = 1.159 T_{rev} = \frac{2\pi + 1}{2\pi}T_{rev}$. The periodic disorder train consists of a repeating series with three spacings of $\tau = 1.159T_{rev}$ and a fourth spacing of $\tau = 1.183T_{rev}$. The aperiodic disorder train is similar to the periodic sequence, except every fourth spacing is a randomly generated value within the range $\tau = [1.159T_{rev}, 1.207T_{rev}]$. The full disordered train uses random values within the range $\tau = [1.159T_{rev}, 1.207T_{rev}]$ for all spacings. The train that excites $^{14}$N with kicks on revival is periodic with all $\tau = T_{rev}$; it is plotted here for reference only.  The rotational energy of $^{14}$N is plotted for comparison to theoretical studies of dynamical localization. The relationship between these two values is not one-to-one but shows matching behaviors, especially in terms of boundedness.  The color maps indicate the J-state population distribution as a function of kick count for each of the applied pulse trains.  The peak intensity used for this simulation is 1.2 $\cdot 10^{14} \frac{W}{cm^{2}}$. $T_{rev}$ is the well known $^{14}$N alignment revival time of 8.38 ps.} \label{disorderComp}
\end{figure*}

We perform a numerical quantum propagation simulation to study the molecular alignment of $^{14}$N using pulse trains consisting of 64 pulses.  A split-step operator method is performed to calculate an effective Hamiltonian in the interaction picture.  The effective Hamiltonian acts upon independent density matrices representing states with different directional angular momentum, $m$.  Step sizes of 1 fs or 10 fs are used depending on whether the electric field is present or not.  The pulse temporal envelope is approximated using a cosine squared pulse profile.  We use the rotational constant $B = 1.998 cm^{-1}$ and centrifugal distortion constant $D = 5.737 \cdot 10^{-6} cm^{-1}$ to simulate the quantum rotor.  We consider multiple extended pulse trains with varying spacing disorder in the temporal spacings and plot the results in fig. \ref{disorderComp}. We display both population alignment $<cos^{2} \theta >_t$, which is the time averaged value of field-free molecular alignment; and the average rotational energy of the sample, $<J^{2} / 2I>$, as functions of the number of pulses applied. The population alignment is an experimentally accessible observable, while the energy is the traditionally analyzed quantity in studies focused on dynamical localization. The results indicate that the two quantities are correlated. A discussion of the relationship between population alignment and energy is presented later in this report.

The result in fig. \ref{disorderComp} shows that the population alignment and energy growth are highly dependent on pulse train periodicity. The population alignment results for the ordered train and the train utilizing a periodic spacing disorder show very limited growth with increasing pulse count. The aperiodic pulse trains with randomly generated pulse spacings do not show clear limitations in growth. These trains yield higher population alignment, meaning that the orientation of $^{14}$N becomes more anisotropic and increasingly favors the z-axis with subsequent kicks. The inset in fig. \ref{disorderComp} shows there is clear evidence that disorded-induced effects are visible with eight pulses at the chosen intensity.

The fully disordered train used in fig. \ref{disorderComp} approaches the highest accessible population alignment value.  The maximum time averaged population alignment for $\delta$-kicked low temperature rotors approaches 0.5.  For finite laser pulses, bandwidth limitations set an upper bound on which $J$ states may be occupied using Raman transitions.  The simulations in fig. \ref{disorderComp} consider states up to $J = 20$ in accord with previous studies \cite{PhysRevLett.113.043002, PhysRevA.88.023426}.  We note that our pulse centered at $800$ nm with $40$ nm bandwidth has an energy cutoff of J $\approx 80$.  Our room temperature gas has a reduced population alignment limit because of significant occupation of states with $m \neq 0$.  The population alignment in our system begins at the isotropic value of $0.33$ and is limited to a maximum value of $\approx 0.40$.


The choice of $\tau$ in fig. \ref{disorderComp} for the periodically kicked rotor is made to be selective of the regime.  Rotational wavepackets made with Raman transitions experience either wavepacket amplification or partial annihilation for many rational values of $\frac{T_{rev}}{\tau}$ \cite{PhysRevA.80.063412, PhysRevA.73.033403, izrailev}, where $T_{rev} = 8.38$ ps, the field-free molecular alignment revival period of $^{14}$N. We avoid these regimes with a proper irrational choice of $\frac{T_{rev}}{\tau}$ for the uniformly spaced pulse train, placing the system in a regime where energy gain is bounded for uniformly spaced trains. Our choice of $\tau = \frac{2\pi + 1}{2\pi}T_{rev}$ is also in adherence to previous work \cite{blumel, PhysRevA.29.1639} that considered first order transitions when using exciting pulses with finite bandwidth. Fig. \ref{disorderComp} includes the $\tau = T_{rev}$ case for comparison. The alignment oscillations for this pulse sequence have been surveyed elsewhere \cite{PhysRevLett.113.043002} and are caused by centrifugal distortion within a real rotor.

The results of the simulation were tested in experiments that introduced kicks with spacing disorder. We create an eight pulse train utilizing a triple nested interferometric pulse stacker to split a single 800 nm, 70 fs pulse originating from a 3 W 1 kHz Ti:Sapph source into eight pulses with estimated peak focal irradiance of $1.2\cdot 10^{14} \frac{W}{cm^2}$ as described previously \cite{PhysRevA.80.063412}. The nested interferometer arms can be adjusted to provide several different patterns of disorder.  Our data shows that four pulses are adequate to initiate dynamical localization.  The pulses are focused into a fused silica cell containing a constant flow of dry nitrogen gas. The kick-induced molecular alignment causes a time dependent optical birefringence in a circularly polarized 400 nm low-intensity overlapping probe pulse with a variable delay. The optical birefringence alters the probe's polarization which is measured using time-gated fast photodiodes placed downstream of a polarizer.  The polarization components are measured for two cases in rapid succession using an optical chopper: with and without the pump beam. The normalized difference is proportional to the molecular alignment for a specified delay time between the pump and probe beams. 

The field free molecular alignment signal is averaged across one revival period after each kick to remove time-dependence and yield population alignment.  ~1400 shots are taken for each time dependent alignment value for each of 125 evenly spaced probe delay values.  Probe delay values for which the kerr effect is present are not considered.  A second delay stage controls the interferometric spacing between the first and second set of four kicks.  The two delay stages constantly cycle across the four choices of $\epsilon$ and the eight revival periods after individual kicks to evenly disperse beam drift effects.  A more detailed discussion of the experimental setup and measurement procedure is in Cryan et al \cite{PhysRevA.80.063412}.

\begin{figure}[h]
\includegraphics[width=0.5\textwidth]{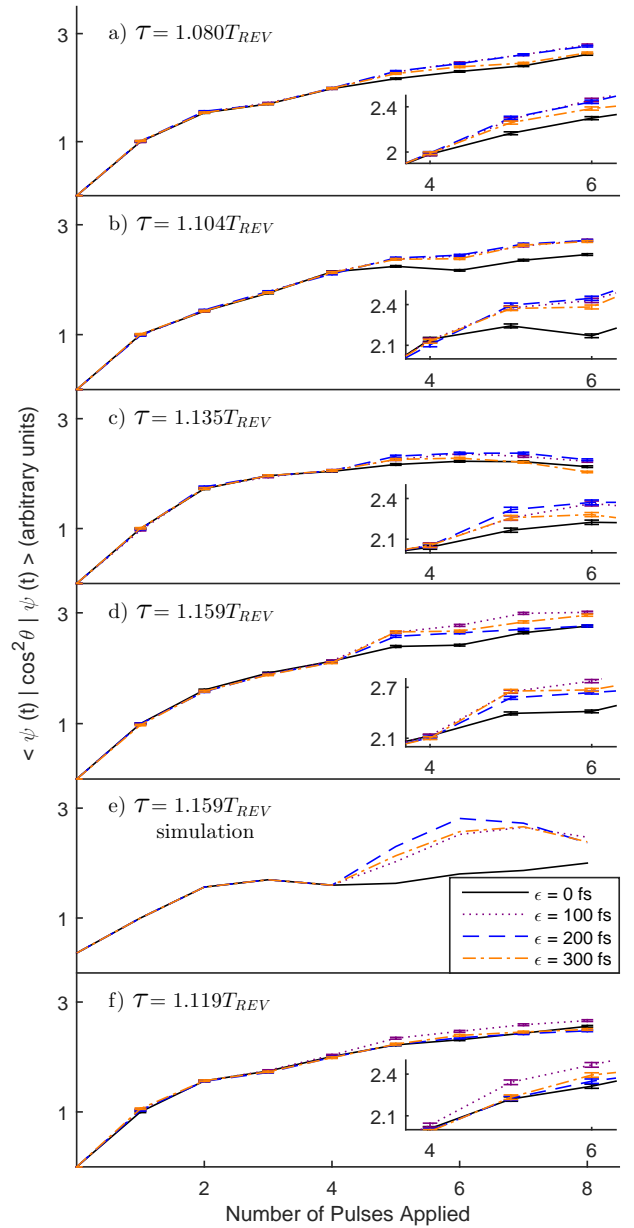}
\caption{The time averaged population alignment of $^{14}$N for experiments studying the effects of spacing disorder. In all cases, the fourth and fifth pulses are separated by a total time of $\tau + \epsilon$ while all other spacings are held constant at $\tau$.  The data in figures $a-d$ shows that fully periodic kicking of the sample causes a minimum gain in population alignment compared to instances of disordered kicking.  Figure $e$ shows the output of the simulation matching experimental parameters in $d$, and the case of $\epsilon = 200 fs$ matches the case of periodic disorder shown in fig. \ref{disorderComp}.  Figure $f$ utilizes $\tau$ for which the fifth pulse overlaps the remaining half revival generated by the initial pulse, causing some wavepacket annihilation.  The intensity is estimated to be approximately $1.2 \cdot 10^{14} \frac{W}{cm^{2}}$. $T_{rev}$ is the well known $^{14}$N alignment revival time of 8.38 ps.}
\label{DLdemoAll}
\end{figure}

We compare the measured population alignment of $^{14}$N excited by an eight pulse train with uniform spacing $\tau$, to $^{14}$N excited by a similar train with a modest amount of time spacing-disorder introduced into the pulse train. The pulse train is set to a value $\tau$ in accord with our previous discussio. The measurement is then repeated after the time separation between the fourth and fifth pulses is changed to $\tau + \epsilon$. A non-zero value of $\epsilon$ thus represents disorder in the pulse spacing. We plot our results in fig. \ref{DLdemoAll}. The results for all measured irrational choices of $\frac{T_{rev}}{\tau}$ show qualitative agreement with simulation results in fig. \ref{disorderComp}, confirming that the introduction of any non-zero $\epsilon$ causes an increase in the population alignment.

Our results are a consequence of dynamical localization within the quantum rotor. The dynamical localization phenomenon corresponds to a predicted localization of rotation state populations within the periodically kicked quantum rotor. We have shown that cases of periodic kicking lead to a slowing of growth in the sample's population alignment in agreement with the predicted onset of localization within the angular momentum state population distribution. Disordered kicking is shown to cause greater change within the population alignment than periodic kicking: this strongly implies that there is greater rotation state population redistribution within the aperiodically kicked rotor, and hence that any localization phenomenon seen for periodic kicking is either absent or significantly weaker in the aperiodically kicked regime. The increased population alignment for disordered kicking means the observed increase in growth within angular momentum space is inherently accompanied by an increase of the molecular orientation along the z-axis.

Simulation results of fig. \ref{disorderComp} also show that cases of disordered but periodic kicking exhibit localized behavior similar to that seen in the strictly periodically kicked rotor. This is not surprising - the link between Anderson localization and dynamical localization \cite{PhysRevLett.49.509} relies on the presence of a periodic Hamiltonian and not necessarily a strictly periodic kicking potential.

The measurement was also performed for a choice of $\tau$ close to a rational fraction of $T_{rev}$. Fig. \ref{DLdemoAll}-f shows data in which ordered kicking does not yield a clear minimum in population alignment. For this set of parameters, the fifth pulse is expected to overlap with the remnants of the recurring half revival created by the first pulse when $\epsilon = 200$ fs. The fifth pulse causes annihilation of the remainder of the initial wavepacket when $\epsilon$ is tuned to the proper spacing, causing a minimum in the population alignment. This rotational wavepacket annihilation was originally demonstrated with two pulses in previous experimental reports \cite{PhysRevA.73.033403}. The data in fig. \ref{DLdemoAll} is taken at $\tau$ that do not risk interference from wavepacket annihilation.

We conclude with a discussion of the relationship between population alignment and angular momentum state populations. The raw data that was used in this work is the molecular alignment signal acquired by measuring the birefringence of a rotational wavepacket as outlined above. The field-free molecular alignment signal is known to have the form
\begin{align}\label{alignSig}
<\psi(t)|cos^{2} \theta |& \psi(t)> = \\
&\sum^{\infty}_{J,m} \alpha^{m}_{J}|a^{m}_{J}|^{2} \nonumber
+ 2 Re[\sum^{\infty}_{J,m}\beta^{m}_{J}a^{m}_{J+2}a^{m*}_{J}e^{i(2J+3)t}]
\end{align}
where $a^{m}_{J}$ is the coefficient of the state $|J,m>$ and the coefficients $\alpha^{m}_{J}$ and $\beta^{m}_{J}$ were initially published by Leibscher \cite{PhysRevA.69.013402}. For our purposes, we reproduce the values
\begin{align}
\alpha^{m}_{J} = \frac{1}{2J+1}[\frac{(J+1)^{2} - m^{2}}{2J+3} + \frac{J^2 - m^2}{2J-1}].
\end{align}
By averaging the molecular alignment signal across $T_{rev}$ between two applied pulses, we discard the second summation term in Eq. \ref{alignSig} and obtain the population alignment value. The values of $\alpha^{m}_{J}$ increase monotonically with increasing $J$ for cases where $m$ is held constant, with the exception of $m = 0$. We note that our simulation and experiment were performed at room temperature, for which only 12$\%$ of the population has $m = 0$. We also note that the pump pulse train is of constant polarization and can change the total angular momentum $J$ of the rotor but keeps $m$, the angular momentum along the $z$-axis constant. The monotonic relationship between $J$ and $\alpha^{m}_{J}$ has the consequence that an overall increase in $J$ within the sample will strongly correlate with an increase in the measured population alignment. This explains why an increase in population alignment strongly implies an upwards redistribution of $J$ and the absence of angular momentum space localization. Similarly, a plateau across a large number of pulses as seen for periodic kicking in fig. \ref{disorderComp} suggests localization within the angular momentum space of the rotor.

This work provides experimental evidence supporting dynamical localization and also shows how disorder may be used to optimize population alignment. This is the first measurement to encompass a broad range of rotation states that demonstrates effects of dynamical localization as the localization regime is approached by making the pump pulse train strictly periodic. One impact of dynamical localization is the ability to selectively excite specific linear rotor isotopes of a sample \cite{PhysRevLett.109.043003, PhysRevA.74.041403}. Within air, for example, periodic kicking of $\tau = T_{rev}$ the revival period of $^{14}$N would excite the J-state distribution within Nitrogen while having a bounded effect on Oxygen due to localization within the angular momentum space of oxygen. We have shown an increase in population alignment is attainable by the introduction of disorder in pulse spacings. A theoretical study \cite{PhysRevA.81.065401} of alignment of SO$_{2}$ shows a similar increase in alignment for disordered kicks, although Pabst's work places all pulse kicks prior to any initial alignment peak. These results prompt further investigation into the effects of non-periodic pulse trains on molecular alignment.

In conclusion, we have studied the effect of non-periodic rotational impulses delivered to a quantum linear rotor. We find that introducing disorder into the spacings between individual pulses of the pulse train results in an increased population alignment of the sample. The increase of population alignment corresponds to an increase in the sample's rotational energy and average angular momentum value, $J$, while conserving the directional angular momentum $m$. The difference in population alignment is explained by the presence of dynamical localization for periodic pulse train kicks, which is no longer present for disordered pulse trains.

We acknowledge Johannes Flo$\ss{}$ and Ilya Averbukh for stimulating discussions on the subject of dynamical localization. James Cryan is thanked for his preliminary work on periodically kicked $^{14}$N, which paved much of the way for the work reported. This research is supported through the Stanford PULSE Institute at the SLAC National Accelerator Laboratory by the U.S. Department of Energy, Office of Basic Energy Sciences.

\vspace{4mm}\noindent $^{\ddagger}$ Corresponding author. \\ akamalov@stanford.edu

\bibliography{biblio}
\bibliographystyle{unsrtnat}

\end{document}